\documentclass[manuscript]{aastex}

\begin{document}

\title{A Comparison of Copper Abundances in Globular Cluster and Halo Field Giant Stars}

\author{Jennifer Simmerer, Christopher Sneden}
\affil{Department of Astronomy and McDonald Observatory, University of Texas, Austin, Texas 78712}
\email{jensim@astro.as.utexas.edu,chris@astro.as.utexas.edu}
\author{Inese I. Ivans}
\affil{Department of Astronomy, California Institute of Technology, Pasadena, CA 91125}
\email{iii@astro.caltech.edu}
\author{Robert P. Kraft}
\affil{UCO/Lick Observatory, University of California, Santa Cruz, CA 95064}
\email{kraft@ucolick.org}
\author{Matthew D. Shetrone} 
\affil{McDonald Observatory, University of Texas, HC 75, Box 1337 L, Fort Davis, TX 79734}
\email{shetrone@shamhat.as.utexas.edu}
\and
\author{Verne V. Smith}
\affil{Department of Physics, University of Texas at El Paso, El Paso, TX 79968}
\email{verne@barium.physics.utep.edu}

\begin{abstract}
	We derive [Cu/Fe] for 117 giant stars in ten globular clusters (M3, M4, M5, M10, M13, M15, M71, NGC 7006, NCG 288, and NGC 362) and find that globular cluster Cu abundances appear to follow [Cu/Fe] trends found in the field. This result is interesting in light of recent work which indicates that the globular cluster $\omega$ Centauri  shows no trend in [Cu/Fe] with [Fe/H] over the abundance range $-2.0 <$ [Fe/H]$ < -0.8$.   Of particular interest are the two clusters M4 and M5.  While at a similar metallicity ([Fe/H] $\sim-1.2$), they differ greatly in some elemental abundances: M4 is largely overabundant in Si, Ba, and La compared to M5.  We find that it is also overabundant in Cu with respect to M5, though this overabundance is in accord with [Cu/Fe] ratios found in the field.  
\end{abstract}

\keywords{Galaxy: abundances---Galaxy: halo---stars: abundances---stars: Population II---globular clusters: general}

\section{Introduction}
 
	Until a large-sample, high resolution study of field stars covering a wide range in metallicity revealed a definite trend in [Cu/Fe] with [Fe/H] \citep{Sneden1988, Sneden1991}, the element copper had not received a  great deal of attention in stellar abundance analyses.  Sneden et al.  found that [Cu/Fe] decreased sharply with [Fe/H], reaching [Cu/Fe]~$\sim -1$ at [Fe/H]~$\sim -3$.  Their reanalysis of other data in the literature  along with averaged globular cluster abundances reinforced their field star findings.  This work has been revisited by \citet{Mishenina2002}, who confirmed this basic result with an expanded sample of halo field giants.

	Despite this clear trend in [Cu/Fe] with [Fe/H] among field stars, interpretation of its origin is distinctly clouded.  In the solar system, copper owes $\sim$30\% of its abundance to the s-process weak (massive star) and main (low mass star) components combined \citep{Matteucci1993}, with the remainder likely a product of some combination of SNe Type Ia and Type II.  Both  the weak and main s-process  are ``secondary'' processes  in the sense that a ``secondary'' element requires a Fe (or Fe-peak) seed nucleus to build upon.  Such an element is expected to decrease  in abundance (relative to Fe) as the star's overall iron content decreases (because the production of the element depends upon the availability of seed Fe nuclei).  In contrast, a ``primary'' process does not require a  Fe seed nucleus, and so the abundance of a primary element (relative to iron) should be constant with overall metallicity.  While [Cu/Fe] does decrease with metallicity in the manner of a  secondary element, the s-process contribution to copper is sharply restricted. Increasing the s-process component of copper in s-process nucleosynthesis models overproduces other  light s-process elements (like Sr), and models of chemical evolution fail to reflect the observed solar composition \citep{Matteucci1993,Baraffe1993}.  The massive star (weak s-process) contribution appears to be an order of magnitude too small to reproduce the Cu content of either the Sun or metal-poor stars, and it must be augmented by explosive nucleosynthesis.  It is unclear whether the additional copper in field stars is produced mainly by Type Ia supernovae  \citep{Matteucci1993, Baraffe1993} or by Type II supernovae \citep{Timmes1995}.

	\citet{Cohen1978, Cohen1979, Cohen1980}'s early observations of copper abundances in giant stars of several globular clusters of varying metallicity ($-2.4 \leq$ [Fe/H] $\leq -0.4$) yielded somewhat ambiguous results.  Cohen found evidence that copper is deficient with respect to iron and that this deficiency may become larger at lower metallicities.  However, uncertainties in the derived abundances related to hyperfine broadening in the Cu line made a clearer delineation of any such trend impossible \citep{Cohen1980}.  Early studies such as these showed no real abundance trends in globular cluster iron-peak elements, copper included (e.g., \citealt{Spite1985}).  

	The situation is further complicated by the behavior of  [Cu/Fe] with respect to [Fe/H] in  $\omega$ Centauri, a massive globular cluster that displays the largest confirmed range in stellar metallicity, with stars as metal poor as [Fe/H]~ $\sim$~-1.8 and as metal rich as [Fe/H]~$\sim$~-0.8 \citep{Norris1996, Suntzeff1996}.  Not only is the [Cu/Fe ]-ratio of $\omega$ Centauri stars uniformly low over  its metallicity range, there is also no evidence of a decrease in [Cu/Fe] with decreasing [Fe/H] \citep{Smith2000, Cunha2002} as is found for field stars over the same interval.  \citet{Cunha2002} suggest that this  could simply be a reflection of $\omega$ Cen's unique chemical history, and may indicate that the cluster built up copper through the contributions of Type II SNe only, contributions from Type Ia SNe being either suppressed  or not retained by the cluster.

	In this paper we present Cu abundances from several other globular clusters spanning the metallicity range $-2.4\lesssim$~[Fe/H]~$\lesssim-0.8$.  Our purpose is to determine whether a large sample of globular cluster stars show the same trend in [Cu/Fe] that field stars do, or whether they follow a pattern similar to $\omega$ Cen.      Cu abundances were derived for bright red giants from M3, M4, M5, M10, M13, M15, M71, NGC~288, NGC~362, and NGC~7006.   A comparison of the [Cu/Fe]-ratios of the two clusters M4 and M5 is especially interesting.  Although the overall metallicity of M4 is very similar to that of M5, it shows significantly higher abundances of the elements Si and Al, as well as overabundances  of s-process  elements like La and Ba compared to M5  \citep{Ivans1999, Ivans2001}.  \citet{Ivans2001} also determined that the abundances of M5 were ``normal'' for population II stars--~that is, M5's [X/Fe] ratios are similar to those found in halo globular cluster and field stars.   

\section{Observations}{\label{obs}}
	All spectra have been previously analyzed by the California/Texas group (hereafter ``CTG''), and we adopt their published model parameters  as summarized in Table \ref{tbl-1}.  In these analyses, the raw spectra were trimmed, flat fielded, extracted, and wavelength calibrated  according to standard procedures using the data analysis packages IRAF\footnote{IRAF is distributed by the National Optical Astronomy Observatories, which are  operated by the Association of Universities for Research in Astronomy, Inc., under cooperative agreement with the National Science Foundation.} and SPECTRE \citep{Fitzpatrick1987}.  For details we refer the reader to the appropriate source (see Table \ref{tbl-1}).  The data set is composed of contributions from several observatories and instruments, and the data quality and wavelength range are not uniform. Spectral resolution ranges from R$\simeq$~30,000 to R$\simeq$~60,000 and S/N is typically $\sim$~70 at the Cu I line at 5782~\AA.    For lower S/N spectra, a two-pixel Gaussian smoothing was applied to the data.  

	The two optical Cu I lines that are strong enough to be useful for abundance determinations in metal poor stars lie at 5782.14~\AA~and 5105.50~\AA.  No detectable ionized lines exist at optical wavelengths.  The spectral region near the 5782~\AA~transition is shown for representative stars from three clusters in Figure \ref{mont}.  This line is strongest in the most metal-rich cluster in the sample (M71) but essentially undetectable in the most metal-poor cluster (M15).  Since wavelength coverage for our spectra is not uniform, the 5105~\AA~line was available only for NGC~288, NGC~362, M4, and those M15 and M13 spectra taken with the  Hamilton echelle spectrograph (at R$\simeq$50,000).  While it is preferable to have independent abundance determinations  based on several different lines, the 5105~\AA~line is likely to be a less reliable Cu abundance indicator in certain cases.    It is an intrinsically stronger transition ($log~gf=-1.52$; \citealt{Bielski1975}) and so comes much closer to saturation as metallicity increases.  This makes the derived abundance more uncertain, since significant changes in the Cu abundance produce only small changes in the appearance of the line.  A strong line also gives an abundance that is more sensitive to the choice of microturbulence.  In contrast, $log~gf=-1.72$ for the 5782~\AA~line, which arises from the same excitation level as the 5105~\AA~line \citep{Bielski1975}.

	Besides these intrinsic problems, the 5105~\AA~line also lies in a more crowded part of the spectrum, especially in the cooler and more metal rich stars.  Several MgH features exist near 5100~\AA~and become quite strong in the cooler stars even at low metallicity, making continuum identification and reliable synthesis fits more difficult.  It is only in the most metal poor cluster studied here, M15 ([Fe/H]$\sim -2.4$), that the MgH lines are absent.  Finally, in these bright but cool giants the spectral energy distribution favors a much higher flux at red wavelengths than at blue wavelengths.  This results in poorer S/N near the  5105~\AA~line relative to the 5782~\AA~line.

\section{\label{ana} Copper Line Analysis}
\subsection{\label{lin}Line parameters}
 	Copper has significant hyperfine components and two stable isotopes: $^{63}$Cu and $^{65}$Cu.  $^{63}$Cu is the more abundant isotope in the solar system, contributing 69\% of the total solar system abundance.   The hyperfine and isotopic structures of the copper lines have significant broadening effects, which become more important as the Cu line gets stronger.  We therefore chose to derive abundances from spectrum synthesis, using  oscillator strengths and hyperfine coefficients given in the Kurucz line list \citep{KurCD1993}. Deriving the Cu abundance from equivalent widths would require that this broadening be somehow accounted for (e.g., artificial addition of microturbulence) with attendant increases in abundance uncertainties \citep{Cohen1978}.  Additionally, in the crowded spectral regions of cool or metal rich stars, a given line is likely to be blended to some extent, resulting in additional uncertainty in an EW analysis.   We therefore used the current version of the LTE line analysis program MOOG \citep{moog} to fit a synthetic spectrum to a 10 \AA~wavelength region around the 5782.14~\AA~Cu I transition.  For those clusters for which we had sufficient wavelength coverage, a 10~\AA~spectrum synthesis was also employed to derive an abundance  from the Cu I line at 5105.5~\AA.  

	The model atmospheres and parameters used in the spectrum synthesis are identical to those derived by the California/Texas group.  We have adopted those previously determined abundances, particularly Mg and C, since the 5105~\AA~region contains a large number of MgH and C$_2$ lines in the cooler and higher metallicity stars.  Though these  molecular lines have been treated as a simple parameter of fit, in only one case has it been necessary to alter either of these abundances.   In the M4 star L2208 the Mg abundance derived from atomic Mg lines by \citet{Ivans1999} did not seem to reproduce  the strong MgH features surrounding the 5105~\AA~line.  In this case, an increase of 0.3 dex to Ivans et al.'s Mg abundance  better fit the observed features.  This star was problematic in the Ivans et al.~study (see  their \S4.2.2 for a discussion of the anomalies found in this star's alpha and light odd-elements.  The data for this star are of lower resolution and lower S/N than most of the sample).  However, since the Mg abundance serves here only as a parameter of the spectrum synthesis fit, it does not affect the Cu abundance we derive.

\subsection{\label{sol}The Solar Cu Abundance}
	   A line list was prepared  by using MOOG \citep{moog} to fit lines taken from the \citet{KurCD1993} line list in a 10\AA~region around each  Cu line to the \citet{Kurucz1984} solar spectrum.  While the fit was in general quite good with the initial line parameters,  $gf$ values for a few Fe lines were modified empirically to produce a better fit.  The Cu I $gf$ and hyperfine values are well determined and were not changed.  Smoothing parameters and the continuum level were  adjusted to minimize the difference between the observed solar spectrum and the computed spectrum.  Once the best fit to the overall spectral region had been achieved, the abundance of Cu was varied to produce the best fit to the Cu I line, and a solar abundance was determined for both Cu I transitions.

	With the Holweger-Mueller solar model ($v_t =1.15$; \citealt{Holweger1974}) we derived $log \epsilon_{\odot}= 4.06$ from the 5782~\AA~line, which agrees well with the value \citet{Cunha2002} derive in a similar manner ($log \epsilon_{\odot}= 4.06$), but is lower than the accepted solar photospheric abundance ($log \epsilon_{\odot}= 4.21 \pm 0.04$; \citealt{Anders1989, Grevesse1998}).  The 5105~\AA~line gives a larger abundance in better agreement with the standard value:  $log \epsilon_{\odot}= 4.21$.  Because of the disagreement, [Cu/Fe]$_{5105}$ is calculated using the 5105~\AA~solar value and [Cu/Fe]$_{5782}$ with the 5782~\AA~solar value.  This is done with the expectation that this will eliminate any systematic errors between the atomic parameters of the lines, and [Cu/Fe] from both lines can be compared without further correction\footnote{As in other CTG work, we have adopted $log~\epsilon_{\odot}~Fe=7.52$ throughout; see \citealt{SKPL1991} for a discussion of this choice.}.  Indeed, [Cu/Fe]$_{5105}$ and [Cu/Fe]$_{5782}$ for cluster stars generally agree when calculated in this manner (see Figure \ref{comp}).

\subsection{Error Analysis}{\label{err}}

	The Cu abundances we derive are sensitive to the adopted stellar parameters T$_{eff}$, $log~g$, $v_t$, and [Fe/H].  We find that a change in T$_{eff}$ of $\pm$ 100 K produces the largest effect in [Cu/H], a change of $\mp$ 0.10 dex. [Cu/H] is less sensitive to the assumed gravity: a change in $log g$ of $\pm$ 0.2 dex produces a change in [Cu/H] of $\pm$ 0.04 dex. However, these changes are largely canceled out in [Cu/Fe] by the accompanying change in [Fe/H] (see, for example, \citealt{Ivans2001}, Table 3).  Factoring in the change in [Fe/H] with T$_{eff}$ and $log~g$ results in $\Delta$ [Cu/Fe] $\pm 0.02$ dex for $\Delta$ T$_{eff} \pm$ 100K  and no discernible change in [Cu/Fe] with a 0.2 dex change in $log~g$.  A change of $\pm$ 0.20 dex in [Fe/H] produces a change of $\pm$ 0.04 in [Cu/H].  The effect of microturbulence varies, being essentially negligible in warm stars to effecting changes of up to $\mp$ 0.10 dex in [Cu/H] for $\Delta v_t = \pm 0.20$ km/s in the coolest stars.  The quality of the synthesis fit to the observed spectrum also varies, depending on S/N.  In high S/N spectra (e.g., M4 and M5), the fit is good to about 0.05 dex.  In other spectra, the fit is good to less than 0.10 dex.  We have therefore adopted an error of 0.10 dex on all our abundances, though in some cases the synthesis fit may be good to only about 0.20 dex (regardless of other error sources).  Those stars with larger fitting uncertainties are marked in Table \ref{tbl-1}.  Additionally, the weak lines in the cluster M15 (while advantageous in some respects)  make abundance determinations more uncertain, and we give all those stars an error of $\pm$ 0.20 dex on the basis of spectrum fitting alone.

\section{CTG Analysis of Globular Cluster Giants}{\label{globana}}
\subsection{Methodology}{\label{meth}}

     Methods of deriving abundances in old metal-poor stars vary considerably from one group of investigators to another. In  \S \ref{globfield} we will compare values of [Cu/Fe] obtained here for globular cluster giants with those obtained by \citet{Mishenina2002} for halo field giants, so we state at this point the assumptions that have gone into the CTG modeling, and a corresponding discussion of the Mishenina et al. methods  can be found in Appendix \ref{ap1}.

     CTG proceeds as follows:

    (1) T$_{eff}$. A preliminary estimate is made from knowledge of the cluster color-magnitude diagram and estimated color excess (e.g., \citealt{Harris1999}),  coupled to a color vs. T$_{eff}$ scale based on (B-V) or preferably (V-K) (e.g., \citealt{Cohen1978}, \citealt{AlonsoGiant}, \citealt{Houdashelt2000}). The preliminary value is modified with the requirement that the value of [Fe/H] derived from the LTE analysis of each Fe I line shall be independent of the line's lower excitation potential. The adopted value of T$_{eff}$ is thus essentially independent of errors in color or color excess.

     (2) $log~g$. It is assumed that LTE is satisfied and that $log \epsilon$(Fe) from Fe II can be brought into agreement with $log \epsilon$(Fe) from Fe I.  This technique avoids uncertainties in the distance scale associated with globular clusters (see, e.g., \citealt{Reid1997, Reid1998} and \citealt{Carretta2000}).  However, \citet{Thevenin1999} have pointed out that NLTE effects in stellar atmospheres lead to the over-ionization of iron, such that measurements of Fe I and Fe II need not result in the same Fe abundance.  This effect has implications for the derivation of both $log~g$ and [Fe/H].  

	Since both [Cu/H] and [Cu/Fe] are insensitive to uncertainties in $log~g$ (see \S \ref{err}), an LTE derivation of this quantity will not affect those results.  The remaining issue, the metallicity scale for these clusters, is beyond the scope of this paper but has been addressed by \citet{Kraft2002}.  \citet{Kraft2002} follow the reasoning of \citet{Thevenin1999} and \citet{Ivans2001}, who argue that Fe II is a better measure of stellar iron abundances than Fe I (because most of the iron atoms in the star are ionized), and measure globular cluster metallicities accordingly.  We have preserved the original LTE metallicities as measured by CTG, under the assumption that the effects of over-ionization on Cu I in a star's atmosphere are somewhat mitigated if one measures abundance ratios of elements with similar ionization potentials in the same ionization state (i.e.,  deriving [Cu/Fe]$_I$; see \citealt{Kraft2002} for a discussion of these issues).   In an LTE analysis, the reported metallicity is essentially the Fe I abundance and our [Fe/H] values differ from \citet{Kraft2002} by varying amounts.  However, maintaining the LTE [Fe/H] scale preserves our ability to directly compare the run of [Cu/Fe]$_I$ with [Fe/H] in clusters with that of the field and $\omega$ Cen, since those studies have also employed the standard LTE analysis.  NLTE effects will distort the [Fe/H] axis of our trends, but legitimate comparisons to field star trends can still be made  so long as we take care to match cluster giant stars  with field giant stars.

     (3) Values of $log~gf$. For Fe I, the adopted values are consistent with the Oxford group \citep{Blackwell1982} and with \citet{Gurtovenko1989} (see also \citealt{Fuhr1988}). For Fe II, CTG adopted those of \citet{Moity1983}, \citet{Heise1990} and \citet{Biemont1991}. A more nearly complete discussion is given in \citet{Sneden1991}.  In M4 and M5, these lines were supplemented by revised $log~gf$'s from \citet{Lambert1996}.

     (4) Models. CTG employs MARCS models for old, metal-poor giants (Gustafsson et al., 1975).  In connection with this exposition of method, we make a special comment on the CTG analysis of giants in M5 and M4. In the M5 paper \citet{Ivans2001} prefer to adopt a procedure in which T$_{eff}$ is derived  from reddening corrected colors and the Alonso et al. $(V-K)_0$ vs. T$_{eff}$ scale, and $log~g$ is derived from knowledge of the cluster distance modulus, this in response to concerns about the reliability of the assumption of LTE (compare \citealt{Thevenin1999}). This alters somewhat the derived abundance of Fe, in comparison with the (above) ``traditional'' method.  This approach is, however, inconsistent with the CTG analysis of the other clusters. Fortunately, Ivans et al. also give the result of the traditional analysis (see Appendix A, Table 11), and we adopt here the value of [Fe/H]$_{avg} = -1.35$ given in the appendix. For M4, which is discussed in the M5 paper in connection with this problem, we also adopt here the metallicity derived from the ``traditional'' approach: [Fe/H]$_{avg} = -1.18$.

     Model parameters, values of [Fe/H] taken from CTG, and values of [Cu/Fe] derived here for individual cluster stars in M71, M4, M5, NGC 362, NGC 288, M3, M10, NGC 7006, M13 and M15 are given in Table \ref{tbl-1}. Cu abundances are normalized to the mean of [Fe/H] derived from Fe I and Fe II.   Some clusters give rise to specific concerns, and we address those in the following section.

\subsection{Special considerations }{\label{special}}

\subsubsection{M4}{\label{M4}}

	M4 was one of the few clusters  in our sample for which both Cu lines were available for analysis.  However, the 5782 \AA~ feature falls close to the end of the spectral orders in this cluster, and so $\sim$ 4\AA~ of the spectrum used for fitting in the other clusters  is unavailable.  This is not likely to be a source of significant error since the spectrum in these stars is very clean (high S/N and relatively weak features).

	A potentially more serious problem in M4 is the presence of a diffuse interstellar band (DIB) at 5780\AA~ (e.g. \citealt{Herbig1975, Krelowski1993}).  Fortunately, the Cu feature is shifted enough by M4's radial velocity that the DIB does not directly affect it (see Figure \ref{m4m5}).  To improve the fit of the synthetic spectrum, the DIB was divided out manually using SPECTRE \citep{Fitzpatrick1987}, effectively treating the feature as continuum.  Comparison of [Cu/Fe]$_{5782}$ to [Cu/Fe]$_{5105}$ indicates that the DIB did not in fact affect the Cu feature and its removal by this simple means has not introduced any appreciable fitting error; the spectrum fits for this line are as good as $\pm$0.05 dex.  In every case but  two (stars L2208 and L4201), the abundances  from the 5782~\AA~and 5105~\AA~lines agree to 0.1 dex.

	Though the 5105 \AA~ line was available for all M4 stars, it was not possible in every case to determine an abundance from it.  In stars cooler than 4225 K (with the exception of L2406) the spectrum was too crowded by strong MgH lines to achieve a reliable fit of the synthesis to the observed spectrum.

\subsubsection{\label{M15} M15}

	M15 is the most metal poor cluster in this sample.  The overall metallicity is so low that the 5782 \AA~ feature is too weak to synthesize even in the cooler stars; all that could be derived from it were unenlightening upper limits.  We have therefore synthesized  only the 5105 \AA~ line in M15 stars.  Given the usually good agreement between the two lines in other clusters, we are confident that the derived abundance from this line accurately reflects the Cu content of a particular star.  However, even the stronger 5105~\AA~line is still quite weak in the M15 spectra, and individual abundances are still uncertain.  We  put larger errors on M15 abundances, and our average abundance for this cluster has a high $\sigma$.

\subsubsection{\label{M71} M71}
	The most metal rich cluster in the sample, M71, posed a different sort of problem for our analysis.  The spectrum becomes so crowded with lines that it is very difficult to determine the level of the continuum.  Only six stars from \citet{Sneden1994} yielded credible results.  In the coolest stars, we found that [Cu/Fe] correlated with T$_{eff}$, with the coolest stars giving the lowest [Cu/Fe].  This is likely due to errors in the continuum fitting.  In the coolest giants, molecular band heads  of TiO are particularly strong (see \citealt{Sneden1994}, their Fig. 1), and a TiO band head does exist near the 5782~\AA~feature (at 5758~\AA, degraded to the red; \citealt{Pearse1963}).  If these band heads are not properly taken into account, the tendency is to incorrectly identify the continuum so that the Cu line becomes shallower than it ought to be and gives a falsely low Cu abundance.  

\section{\label{globcu} [Cu/Fe] Ratios in Globular Clusters}
	Copper lies at the end of the Fe-peak and may or may not actually be  formed in the same processes that create the lighter elements of the Fe-peak.  However, like the  Fe-peak elements, it shows no real star to star variation in these globular clusters.  The standard deviation of the mean [Cu/Fe] value for every cluster in the sample is quite small and well within the observational errors.  In Table \ref{tbl-2} we list for each cluster the values [Fe/H]$_{avg}$,  [Cu/Fe]$_{avg}$, and the standard deviation ($\sigma$), and in Figure \ref{trend}, we plot [Cu/Fe]$_{avg}$ vs. [Fe/H]$_{avg}$ for the clusters in our sample.

	As a supplement to Figure \ref{trend}, we offer in Figure \ref{box} box plots in which the clusters are arranged in order of  increasing metallicity, and which illustrate both the median [Cu/Fe] and the range in [Cu/Fe] for each cluster.  Figure \ref{box} appears to show a rather large range in [Cu/Fe] in M10, when the outliers are considered.  However, an examination of the spread in [Ni/Fe] (which ought not to vary from star to star) in M10 as measured by \citet{Kraft1995} indicates that the spread in [Cu/Fe] is not real: an investigation of each of the five Ni lines analyzed by \citet{Kraft1995} reveals that $\sigma$([Ni/Fe]) ranges from 0.10 dex to 0.15 dex.  $\sigma$([Cu/Fe])=0.10 is therefore not significant.  The total range of [Ni/Fe] in each individual Ni line is as much as 0.3 dex in this cluster.  It should also be noted that the single low outlying [Cu/Fe] value, M10:G, is the warmest star in our M10 sample, and has a very weak line and low S/N.  \citet{Kraft1995} were able to measure only one Ni line in that star.  In M4, a cluster for which we have a large number of high quality spectra, $\sigma$([Ni/Fe]) is also comparable to $\sigma$([Cu/Fe]), which suggests that the spread in [Cu/Fe] in a particular cluster is related to data quality rather than any actual star to star variation.

     Among halo field stars, which are generally regarded as surrogates of globular cluster stars, \citet{Sneden1991} found a sharp decrease  in [Cu/Fe] with decreasing metallicity (a mean of Fe-peak elements including Ni but dominated by Fe), following the mean relationship [Cu/M]$ = +0.38[M/H] + 0.15$. We plot this also in Figure \ref{trend}.  It is seen that in general the clusters follow a fairly similar decline with roughly the same slope as that of the field stars, but displaced toward slightly higher values of [Fe/H] and/or lower values of [Cu/Fe]. With the exception of M15, all clusters lie ``below'' the line given by \citet{Sneden1991}. 
	
      The displacement between field and cluster stars noted from inspection of Figure  \ref{trend} may possibly be real, but may also be a result of systematic differences between investigators. We explore this possibility further using the expanded sample of halo field stars recently studied by \citet{Mishenina2002}, who explored [Cu/Fe]-ratios in a sample of 90 metal-poor stars in the range $-3.0 <$ [Fe/H] $< -0.5$, on the basis of high-resolution, high S/N spectra.

 \section{Clusters  vs. Field}{\label{globfield}

     Out of the 90-star sample of Mishenina et al., we first selected for comparison with cluster giants the 21 low metallicity ([Fe/H] $< -0.50$) field giants having $log~g \leq $ 2.3. These are essentially the field star surrogates of the cluster giants; elimination of the others suppresses any possible systematic effects owing to a large shift in $log~g$ and those NLTE effects discussed in \S \ref{meth}. In addition, the lower luminosity sample is heavily weighted toward stars having metallicities [Fe/H] $> -1.0$, a regime more nearly associated with the thick disk than with our (predominantly) halo globular clusters.

     Before making the comparison of [Cu/Fe] values among these 21 giants with our globular cluster giants, we must first eliminate as far as possible systematic effects.  The details of their abundance method and the resulting offsets from this work are described in Appendix \ref{ap1}, and those results are listed in Table \ref{tbl-2}.

     In Figure \ref{mishcu} we plot [Cu/Fe] vs. [Fe/H] for both clusters and halo field giants, having adjusted the clusters to system of abundances defined by Mishenina et al. (Obviously, we could have normalized equally well to the system defined by CTG.) The shifts required to move all stars to the same abundance system are small: this is largely because the methods of analysis are quite similar. Inspection of Fig 6 shows that the cluster and halo field giants have a nearly identical distribution of [Cu/Fe] as a function of [Fe/H]. From [Fe/H] near  $-0.7$, the value of [Cu/Fe] is near $-0.1$, then it declines to about $-0.65$ at [Fe/H]$\sim -1.7$, after which it levels off at a constant value near $-0.65$ as [Fe/H] approaches  $-3.0$. 

\section{\label{con} Conclusion} 
\subsection{M4 and M5}{\label{M45}}
	
	The two clusters M4 and M5, while very similar in [Fe/H], show a significant difference in their copper abundance.  This is not likely to be a result of error in the analysis or in stellar atmospheric parameters, given the very small range in [Cu/Fe] derived for these clusters over a large range in T$_{eff}$ and $log~g$.  Furthermore, the difference in the copper abundance between similar stars from different clusters reinforces the difference seen on average. Figure \ref{excess} illustrates this with L2206 and IV-34, stars typical of the M4 and M5 data sets.  These two stars have very similar atmospheric parameters, yet a simple examination of their spectra shows that Cu is indeed overabundant in M4.  While (with one exception) no other species in this region shows significant line depth contrasts, the flux at the 5782~\AA~line in the M4 star is $\sim$1.3 times weaker than in the M5 star.  Comparing this line with others, it is evident that the effect is much greater than simple noise or continuum uncertainties.  The other excess present in Figure \ref{excess} is Si, an element shown by \citet{Ivans1999} to be overabundant in M4.  Unlike other elements in M4, however, Cu does not appear to be unusually enhanced compared to field halo stars.  In this metallicity region, the field halo stars show a very swift increase in [Cu/Fe] that appears to be mirrored in M4, and continues in the higher metallicity clusters M71 just as it does in the field.  Whatever M4's source of enrichment may be that sets it apart in other elements, it has not significantly contributed to the Cu abundance (any more than is ``normal'' for its [Fe/H]).  

	Like M4/M5, NGC~288/NGC~362 have similar metallicities.  NGC~288 and M4 have similar Na-O and Al-O anticorrelations while M5 and NGC~362 have similar Na-O and Al-O anticorrelations \citep{Shetrone2000}.  Unlike the M4/M5 pair, the NGC~288/NGC~32 pair does not show such a large discrepancy in [Cu/Fe].  Though NGC~288 is elevated in Cu, this effect is smaller than in M4, and the uncertainty in NGC 288's Cu abundance  ($\sigma=0.11$) is large.  Though NGC 362's Cu abundance is relatively well determined, that of NGC 288 is not, and we can draw no conclusions as to  Cu abundance differences between the two. These two clusters also seem to follow the field halo star Cu abundance trend, though M4, M5, NGC 288, and NGC 362 all fall near the [Fe/H] value where the trend in [Cu/Fe] with [Fe/H] changes shape  in the field (see Fig. \ref{mishcu}).  

\subsection{Clusters vs. $\omega$ Cen}

	Cluster and field halo giants both display a trend in [Cu/Fe] with [Fe/H] that is totally absent in $\omega$ Cen, where [Cu/Fe] is uniformly deficient over a wide range in [Fe/H].  At or about the metallicity of M5 ($\simeq -1.35$), [Cu/Fe] begins to increase very quickly with [Fe/H] in the field and in the higher metallicity clusters. 

	\citet{Ivans1999} found overabundances of $s$-process elements like Ba and La in M4.  $\omega$ Cen also shows evidence of $s$-process enhancement in its more metal-rich stars \citep{Cunha2002}.  It would be tempting to attribute M4's increased Cu abundance (and therefore also that of the field) to enhanced $s$-processing as well.  However, the bulk of solar system Cu is made from the weak, not main, $s$-process, unlike Ba and La.  Only about 10\% of the solar system Cu can be attributed to the main $s$-process \citep{Kappeler1989}.  \cite{Baraffe1993} find that the massive star (weak $s$-process) contribution is an order of magnitude too small (for all [Fe/H]) to reproduce the field star abundance, and increasing the main $s$-process contribution to M4 should not make much difference in the Cu abundance.  In any case, even $s$-process enhanced stars in $\omega$ Cen still show a low [Cu/Fe] (though there is some Cu enhancement of giants at the highest metallicities; see \citealt{Pancino2002}).    \citet{Cunha2002} attribute the low [Cu/Fe] in $\omega$ Cen to enhanced SNe Type II (and little SNe Type Ia) enrichment.  If so, then the enhancement in M4's Cu  cannot be due to Type II SNe, either.  This conclusion is reinforced by the fact that the Cu  abundances in M15 stars do not correlate with the Eu abundances.  Eu, an $r$-process element, is commonly associated with Type II SNe.  If Cu owed its production to the same process, then it ought to show the same star-to-star variation seen in the M15 Eu abundances.  An inspection of the Eu abundances  in \citet{Sneden1997} for M15 stars shows that the highest and lowest [Eu/Fe] stars in our sample, K462 and K583, have virtually identical [Cu/Fe].  The highest and lowest [Cu/Fe] stars, K634 and K853, have  essentially the same [Eu/Fe].  There is therefore no indication that [Cu/Fe] is in any way related to the SNe Type II tracer  [Eu/Fe] in M15.     

	 Taken together, the Cu abundances in globular clusters and halo field stars suggest a third possibility, that at higher metallicity stellar Cu abundances receive a significant contribution from Type Ia SNe.  The abundance pattern of $\omega$ Cen, where [Cu/Fe] is low, is likely the product of Type II SNe \citep{Smith2000, Cunha2002}, making it improbable that Cu comes from this source, as \citet{Timmes1995} suggest (though Timmes et al. do show that the copper yield of SNeII is metallicity-dependent).  Substantial $s$-process  contributions are ruled out on the basis that M4, a cluster known to be $s$-process enhanced, shows no evidence for an unusual Cu enhancement.  \citet{Matteucci1993} have suggested that Cu may be produced by SNe Ia, a source that seems increasingly likely.  Possibly the best way to resolve this question is to further study M4 and M5's abundance patterns; these two clusters lie at a metallicity where the nucleosynthetic source of Cu appears to undergo a transition.  In particular, the relative abundances of Zn (the next heavier element) in these two clusters may shed some light on the origin of Cu.
 
\acknowledgments
	We thank the referee for providing helpful comments.  We also thank our colleagues who contributed to the original papers that provided the data for this study. We gratefully acknowledge T. Mishenina and collaborators for sharing their results in advance of publication.  We profited from helpful discussions with David Lambert, Peter Hoeflich, and David Yong.  This work has been supported by several NSF grants, most recently by AST 9987162 to CS.  JS is pleased to acknowledge financial support from The University of Texas Austin Graduate Research Internship Program.  III gratefully acknowledges financial support of Continuing University Fellowships from The University of Texas at Austin.
	Some of the data presented herein were obtained at the W.M. Keck
Observatory, which is operated as a scientific partnership among the California Institute of Technology, the University of California and the National Aeronautics and Space Administration. The Observatory was made possible by the generous financial support of the W.M. Keck Foundation.  This research has made use of NASA's Astrophysics Data System.

\appendix
\section{\label{ap1} Appendix}

In order to determine systematic offsets between our cluster analysis and the field star analysis by \citet{Mishenina2002}, we take up the procedures employed by Mishenina et al. and invite the reader to compare these with the CTG methods described in \S \ref{meth}. The actual stellar models, however, are not discussed in \citet{Mishenina2002}, but rather are adopted from an earlier paper, \citet{Mishenina2001}. 

     (1) Equivalent widths. These are in excellent agreement with those obtained by \citet{Pilachowski1996} (see \citealt{Mishenina2001}, their Fig 3), which are on the system of CTG.

     (2) T$_{eff}$. The models are assigned preliminary values of T$_{eff}$ from colors and the Alonso et al. scale of color vs. T$_{eff}$. However, final assignment of T$_{eff}$ depends upon achieving ``flatness'' in the Fe I excitation plot. The procedure for setting T$_{eff}$ in the end is identical to that of CTG.

     (3) $log~g$. The abundance of Fe from Fe II is brought into agreement with that of Fe from Fe I to set $log~g$, again a procedure identical with that adopted by CTG.

     (4) $log~gf$ values. For Fe I and Fe II, Mishenina et al. adopt the $log~gf$ values (as do Mishenina \& Kovtyukh) of \citet{Gurtovenko1989}. For the 30 Fe I lines in common between CTG and Mishenina \& Kovtyukh, we find the remarkable result that $\Delta~log~gf=0.000 ~(\sigma=0.066)$.  For the six Fe II lines employed by CTG, there is a small but significant offset, $\Delta~log~gf=+0.108 ~(\sigma= 0.064)$, in the sense CTG minus Mishenina \& Kovtyukh. We may therefore expect to find a difference in the values of $log~g$ between the present investigation and that of Mishenina et al.

       For Cu, Mishenina et al. adopted the $log~gf$ values given by \citet{Gurtovenko1989} from inverted solar analysis. For the Cu line at 5782~\AA, their $log~gf$ value is identical with that of CTG. For  the line at 5105~\AA, their adopted $log~gf$ value is smaller than that of CTG by 0.06 dex. Thus from this line alone, they  would derive a larger abundance of Cu than we derive in the present investigation; this amounts to 0.03 dex, averaging over the two lines. Although we do not apply this ``correction'' in the comparison of results, we note that 5105~\AA~is analyzed in this paper only for a few giants in M4, NGC 362, NGC 288, M13 and all stars in M15.

    (5) Adopted models. Mishenina et al., following Mishenina \& Kovtyukh, adopt Kurucz models with convective overshoot turned on \citep{KurCD1993}. The differences induced in [Fe/H] among metal-poor giants by switching from MARCS to Kurucz models has been explored by Kraft \& Ivans (2002). Among intermediately metal-poor stars ($-1.7~<$~[Fe/H]~$<~-1.2$), the effect on [Fe/H] (from Fe I) is negligible ($\lesssim$ 0.01 dex), but rises to 0.07 to 0.10 dex among relatively metal-rich and metal-poor giants.

     Fortunately, of the 21 halo field giants considered here, 11 have also been studied by CTG \citep{Kraft1992, Shetrone1996}. We can then empirically determine the offsets in T$_{eff}$, $log~g$, and [Fe/H] which must be applied to bring the field and cluster giants onto the same abundance system. In the sense CTG minus Mishenina \& Kovtyukh, we find the following mean differences: $\Delta$~T$_{eff}=+47K ~(\sigma=89.5K)$; $\Delta~log~g=+0.28 ~(\sigma=0.20)$; $\Delta$~[Fe/H]$=+0.05 ~(\sigma=0.10)$.

     The offsets in T$_{eff}$ and [Fe/H] are small (if significant), but the offset in $log~g$ is rather large. Fortunately, the values of [Cu/Fe] are not very sensitive to variations in $log~g$. In addition, and more importantly, the offset is just about what one would expect owing to the differences in the $log~gf$ values for Fe II. If [Fe/H] from Fe II is brought into agreement with [Fe/H] from Fe I, then a shift of 0.11 in the Fe II $log~gf$ values corresponds to a shift in $log~g$ of 0.20 dex (Ivans et al. 2001, Table 3), close to the value of 0.28 cited above. The [Fe/H] offset cited above is also explicable in terms of the offset in T$_{eff}$. Again, according to the Ivans et al. Table 3, an increase in T$_{eff}$ of 47K should generate an increase of 0.04 dex in [Fe/H] from Fe I, almost exactly that observed.

      Having made the appropriate adjustments in [Fe/H] and T$_{eff}$, it remains only to adjust the values of [Fe/H] from one set of atmospheric models to the other. In columns 5 and 6 of Table \ref{tbl-2}, we list also mean cluster values of [Fe/H] and [Cu/Fe] on the basis of Kurucz models. There is almost no effect for most of the clusters, but there are significant shifts in the case of M71 and M15.

\clearpage

\begin{figure}
\plotone{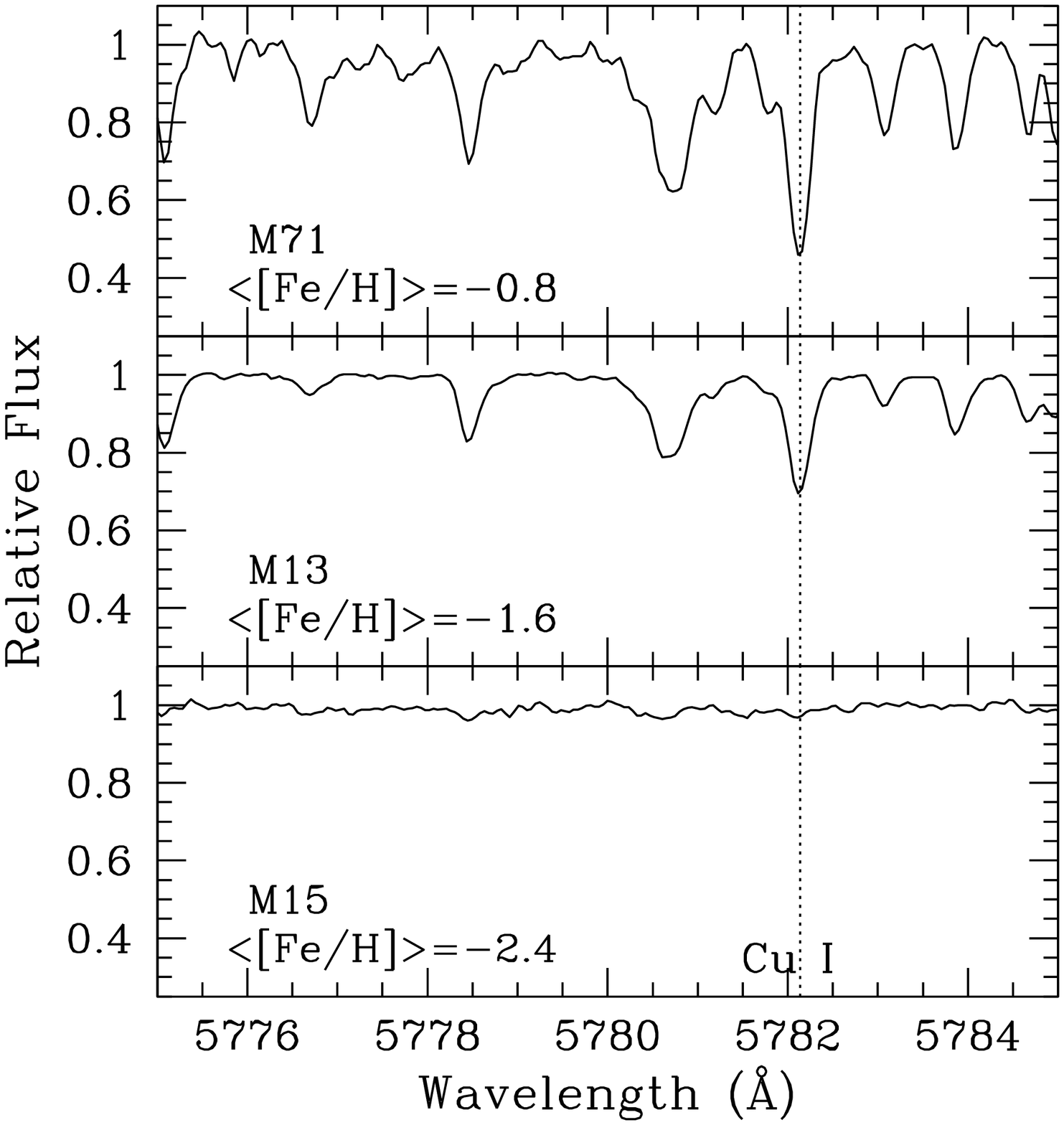}
\caption{Representative Cu I 5782~\AA~lines in stars of three clusters: M71 (upper panel), M13 (middle  panel), and M15 (bottom panel).  The 5782.14~\AA~line is marked in each.  M71 and M15 are the metallicity extremes in the sample.  Most of the clusters have [Fe/H] similar to M13. \label{mont}}
\end{figure}

\begin{figure}
\plotone{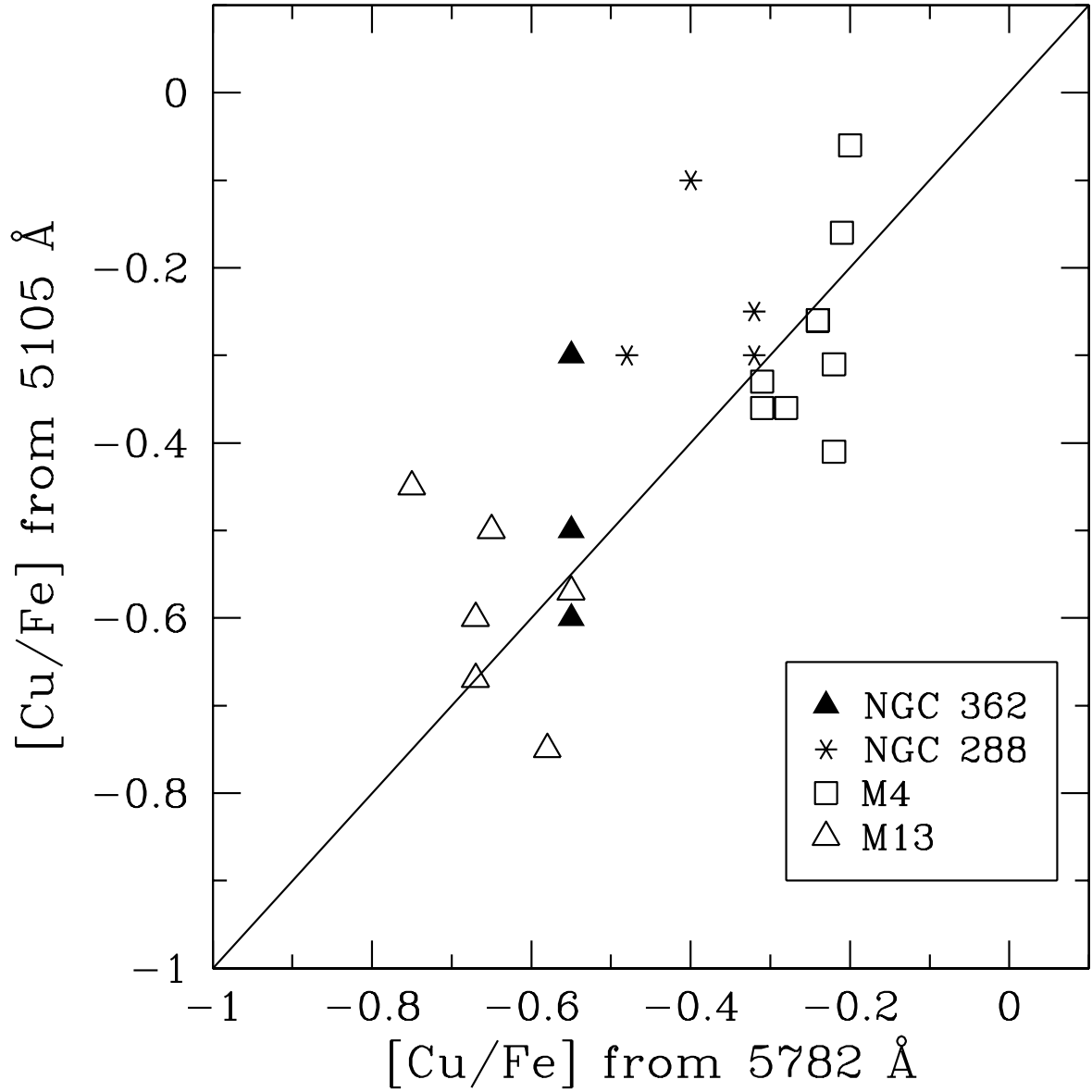}
\caption{Comparison of copper abundances derived from the 5782\AA~ and 5105\AA~ lines. In most cases abundances from the two lines agree to within 0.10 dex.  \label{comp}}
\end{figure}

\clearpage 

\begin{figure}
\plotone{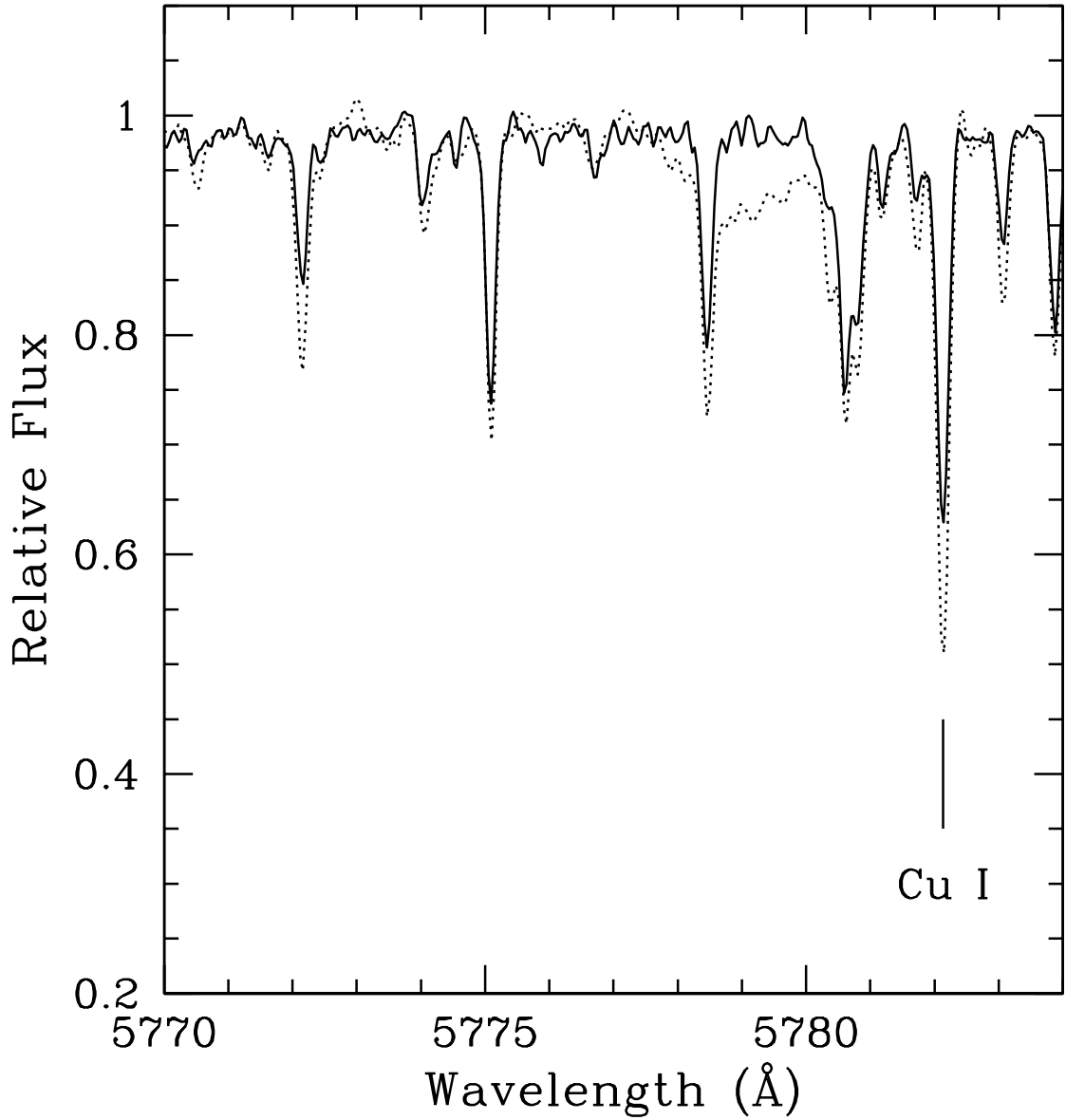}
\caption{Smoothed spectrum of M4:L2206 (dotted line) and M5:IV-34 (solid line).  The stars have similar  T$_{eff}$, $log~g$, [Fe/H], and microturbulence.  The dip in the M4 star's continuum is due to a DIB near 5780~\AA, though it clearly does not extend into the Cu I feature. \label{m4m5}}
\end{figure}
\clearpage 

\clearpage 
\begin{figure}
\plotone{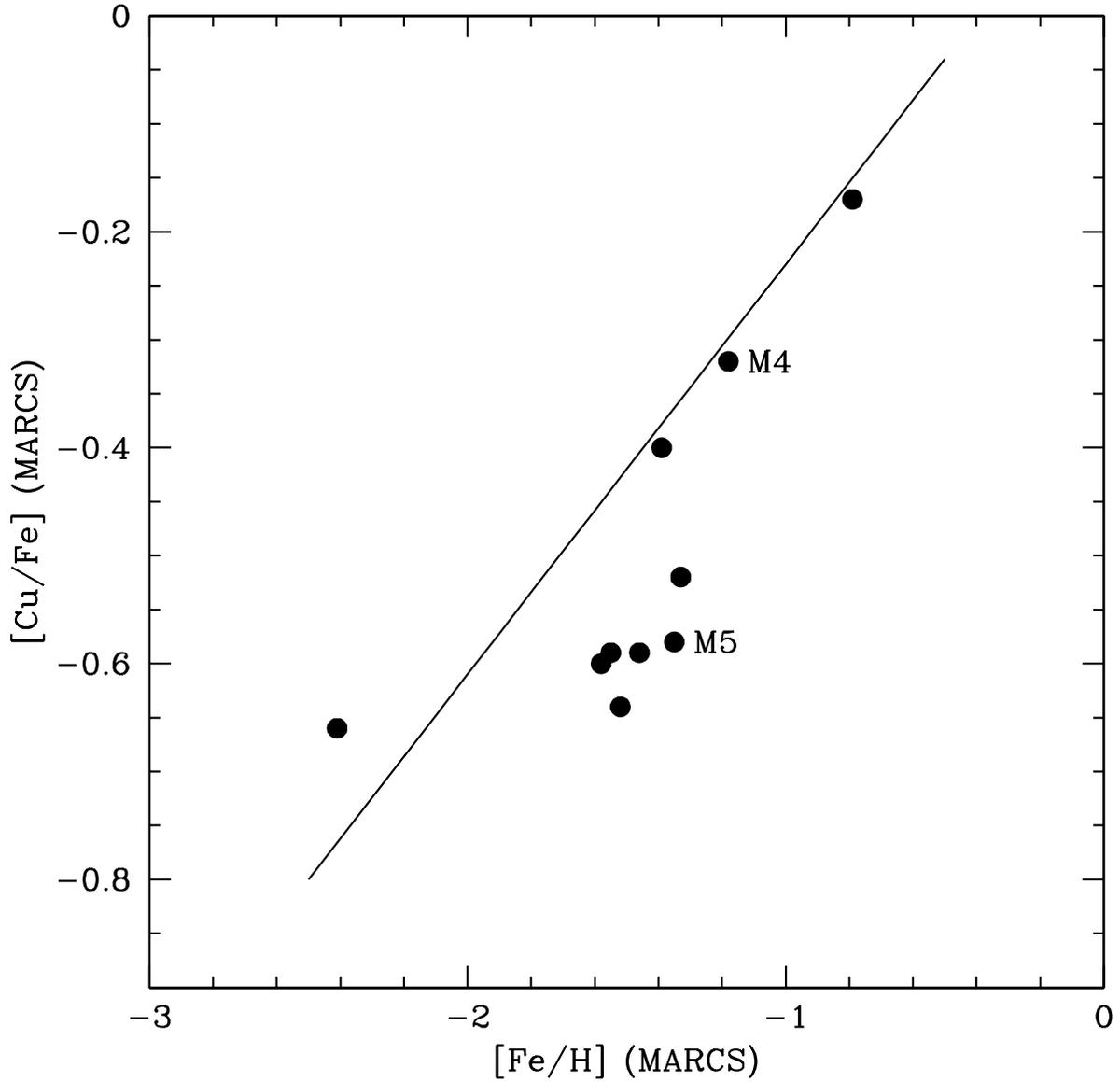}
\caption{Average cluster [Cu/Fe] values on the MARCS model scale.  The line indicates the trend found by \citet{Sneden1991}.  Most clusters fall below the trend, but the overall shape appears similar.  \label{trend}}
\end{figure}

\clearpage
\begin{figure}
\epsscale{0.8}
%%\rotatebox{90}
\plotone{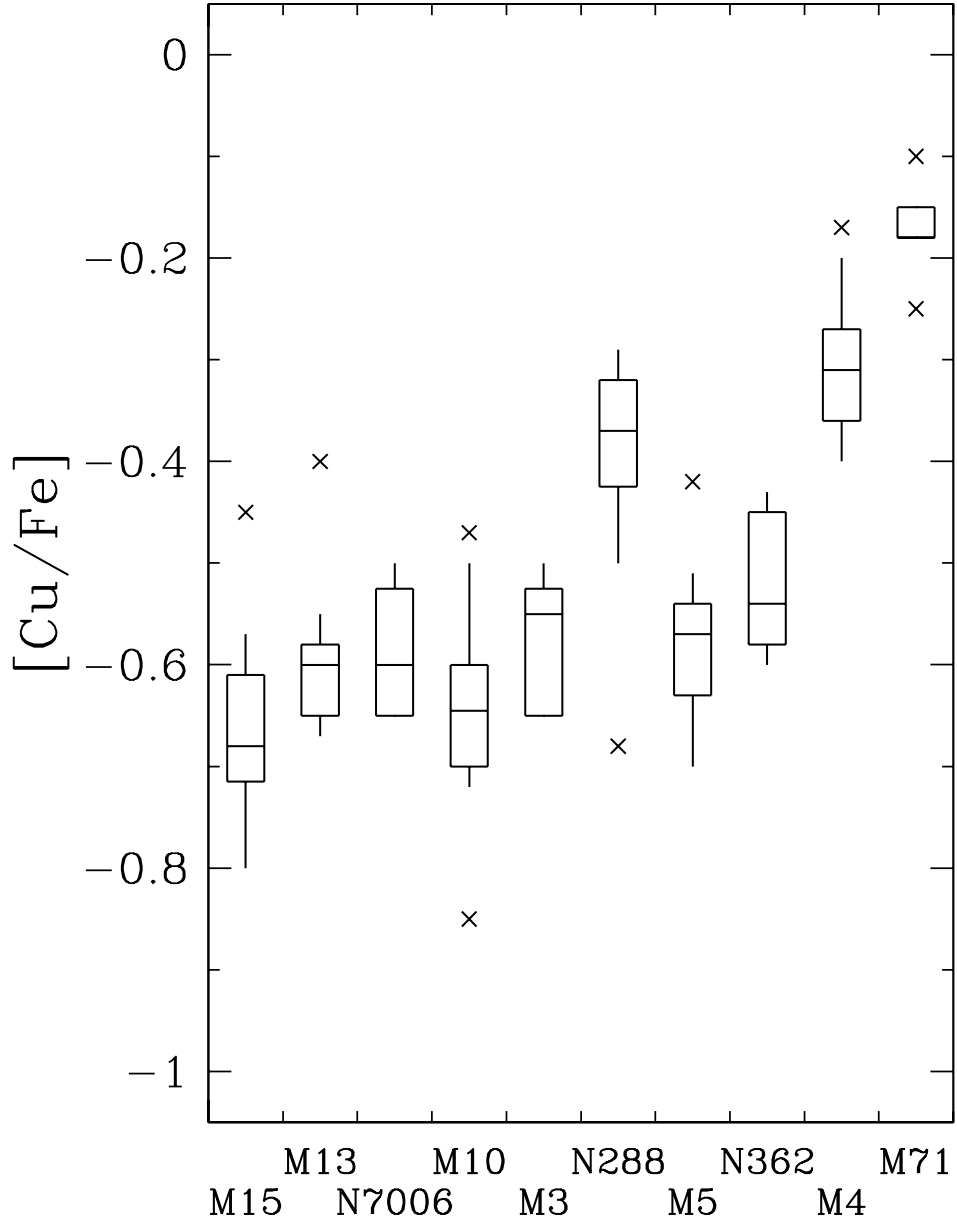}
\caption{The boxed region for each cluster encompasses the interquartile range (middle 50\%) of the [Cu/Fe] data for that cluster.  Also  shown are the median (horizontal line), range (vertical lines; excludes outliers) and outliers (crosses; an outlier has a value greater than 1.5 time the interquartile range).  The clusters are arranged in order of increasing [Fe/H]$_{avg}$.  These values are not corrected to the Mishenina et al. system. \label{box}}
\end{figure}

\clearpage
\begin{figure}
\plotone{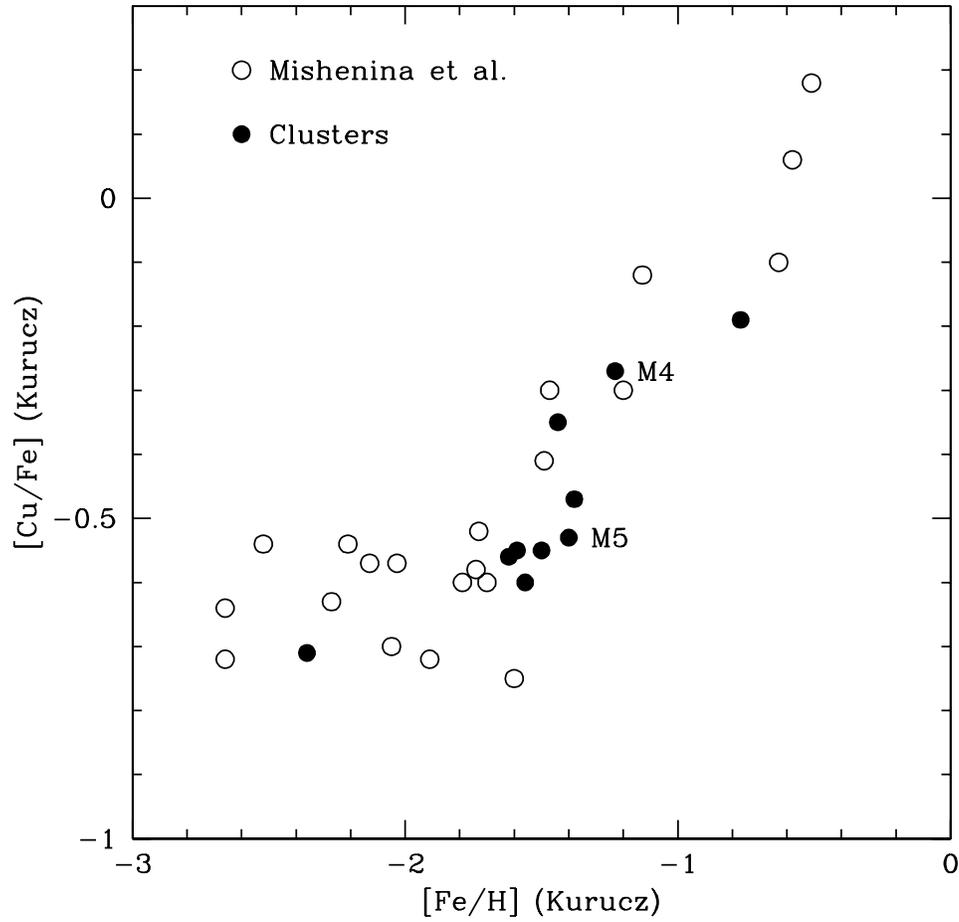}
\caption{Cluster averaged [Cu/Fe] corrected onto the system used by \citet{Mishenina2002}.  The plotted field stars are giants, in keeping with the CTG sample.  There is no indication that the cluster stars deviate from the field field stars in [Cu/Fe] at a comparable [Fe/H].  This also holds true for M4, a cluster that is overabundant in Ba, La, and Si. \label{mishcu}}
\end{figure}

\clearpage 

\clearpage 
\begin{figure}
\plotone{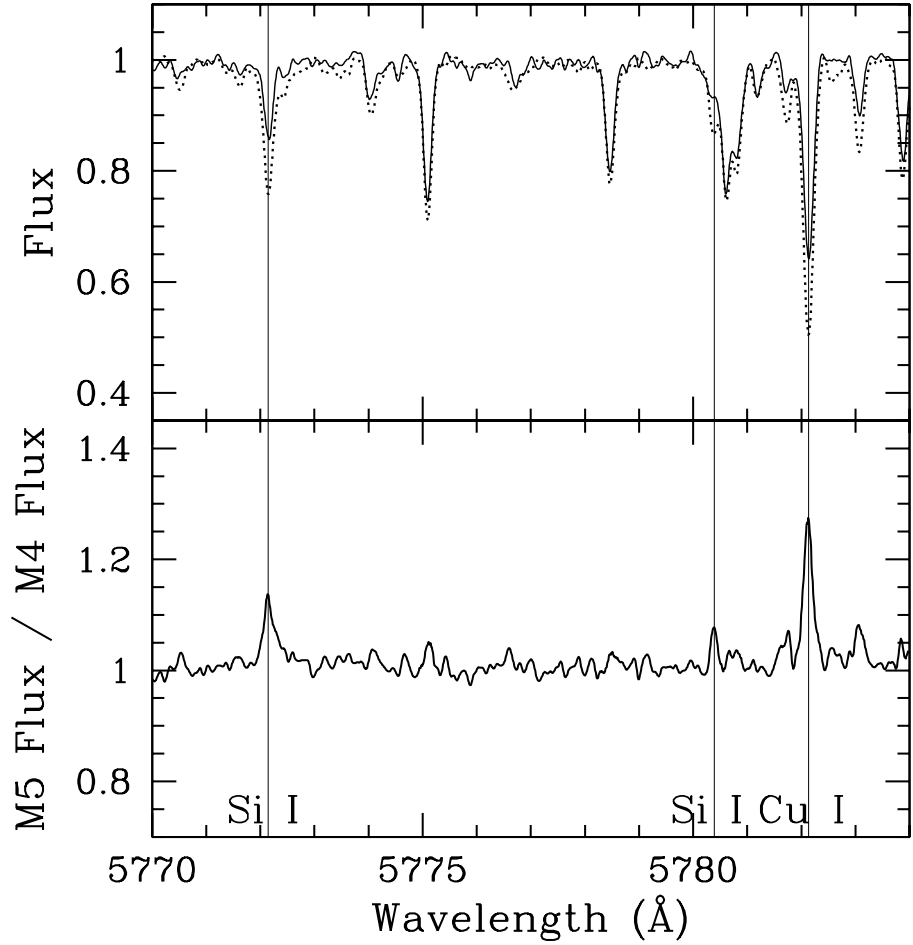}
\caption{Upper panel: Smoothed, interpolated spectrum of M4:L2206 (dotted line) and M5:IV-34 (solid line), two stars with similar T$_{eff}$, $log~g$, [Fe/H], and microturbulence.  Lower panel: the division of the M5 star by the M4 star, illustrating the excess depth of the M4 star's Cu I line. The positions of two Si I lines have also been marked, affirming Ivans et al.'s claim of an excess of Si in M4. \label{excess}}
\end{figure}

\clearpage

\begin{deluxetable}{cccccccc}
\tabletypesize{\scriptsize}
\tablecaption{Model parameters and derived [Cu/Fe] \label{tbl-1}}
\tablewidth{0pt}
\tablehead{
\colhead{Star}& \colhead{T$_{eff}$}   & \colhead{$log~g$}   &
\colhead{$v_t (km/s)$} &
\colhead{[Fe/H]}  & \colhead{[Cu/Fe]$_{5105\AA}$} & \colhead{[Cu/Fe]$_{5782\AA}$} &
\colhead{[Cu/Fe]$_{avg}$}  
}
\startdata
\sidehead{M71 \citep{Sneden1994}}
1-77&4100&0.95&2.00&-0.76&\nodata&-0.25&-0.25\\
A4&4100&0.80&2.25&-0.76&\nodata&-0.18&-0.18\\
A9&4200&1.20&2.00&-0.84&\nodata&-0.15&-0.15\\
1-53&4300&1.40&2.00&-0.82&\nodata&-0.18&-0.18\\
S&4300&1.25&2.00&-0.72&\nodata&-0.15&-0.15\\
01-21&4350&1.45&2.00&-0.74&\nodata&-0.10&-0.10\\
\sidehead{M4  \citep{Ivans1999}}                        
4611&3725&0.30&1.70&-1.16&\nodata&-0.25&-0.25\\
4613&3750&0.20&1.65&-1.17&\nodata&-0.40&-0.40\\
1514&3875&0.35&1.95&-1.16&\nodata&-0.30&-0.30\\
1411&3950&0.60&1.65&-1.20&\nodata&-0.25&-0.25\\
3209&3975&0.65&1.75&-1.20&\nodata&-0.32&-0.32\\
2307&4075&0.85&1.45&-1.19&\nodata&-0.40&-0.40\\
2406&4100&0.45&2.45&-1.20&-0.20&-0.25&-0.23\\
4511&4150&1.10&1.55&-1.18&\nodata&-0.37\tablenotemark{a}&-0.37\tablenotemark{a}\\
3413&4175&1.20&1.65&-1.17&\nodata&-0.27&-0.27\\
2617&4200&0.95&1.55&-1.18&\nodata&-0.35&-0.35\\
3624&4225&1.10&1.45&-1.17&\nodata&-0.32&-0.32\\
3612&4250&1.10&1.45&-1.19&-0.20&\nodata&-0.20\\
2206&4325&1.35&1.55&-1.20&-0.30&-0.28&-0.29\\
2208&4350&1.40&1.70&-1.17&-0.10&-0.24&-0.17\\
4201&4450&1.35&1.85&-1.18&-0.45&-0.26&-0.36\\
1408&4525&1.30&1.70&-1.20&-0.30&-0.28&-0.29\\
1701&4625&1.50&1.65&-1.20&-0.40&-0.32&-0.36\\
3207&4700&1.65&1.70&-1.17&-0.37&-0.35&-0.36\\
3215&4775&1.40&1.85&-1.20&-0.40&-0.35&-0.37\\
4302&4775&1.45&1.80&-1.19&-0.35&-0.26&-0.31\\
\sidehead{NGC 362  \citep{Shetrone2000}}                      
1401&3875&0.00&1.90&-1.32&\nodata&-0.55&-0.55\\
2115&3900&0.00&2.30&-1.38&\nodata&-0.55&-0.55\\
1423&3950&0.10&2.35&-1.37&\nodata&-0.45&-0.45\\
1334&3975&0.40&1.95&-1.30&\nodata&-0.65&-0.65\\
1441&3975&0.20&1.90&-1.31&-0.60&-0.55&-0.57\\
2423&4000&0.40&1.85&-1.32&\nodata&-0.60&-0.60\\
1137&4000&0.70&2.00&-1.37&\nodata&-0.43&-0.43\\
MB2&4100&0.60&2.50&-1.33&\nodata&-0.45&-0.45\\
2127&4110&0.60&2.25&-1.30&-0.30&-0.55&-0.43\\
1159&4125&0.80&1.90&-1.27&-0.50&-0.55&-0.53\\
\sidehead{M5  \citep{Ivans2001}}
G2&3900&-0.10&1.75&-1.35&\nodata&-0.66&-0.66\\
IV-81&3945&0.00&1.90&-1.39&\nodata&-0.58&-0.58\\
I-20&4050&0.00&2.00&-1.47&\nodata&-0.51&-0.51\\
II-85&4050&0.45&1.85&-1.33&\nodata&-0.42&-0.42\\
IV-47&4110&0.50&1.85&-1.33&\nodata&-0.52&-0.52\\
IV-19&4125&0.50&1.70&-1.39&\nodata&-0.51&-0.51\\
III-149&4225&0.60&1.70&-1.33&\nodata&-0.56&-0.56\\
I-14&4250&0.75&1.60&-1.33&\nodata&-0.58&-0.58\\
IV-34&4275&0.65&1.55&-1.32&\nodata&-0.58&-0.58\\
I-58&4350&0.80&1.50&-1.31&\nodata&-0.56&-0.56\\
I-61&4400&1.00&1.50&-1.30&\nodata&-0.57&-0.57\\
II-59&4463&1.15&1.65&-1.25&\nodata&-0.61\tablenotemark{a}&-0.61\tablenotemark{a}\\
III-18&4475&0.55&1.70&-1.46&\nodata&-0.59&-0.59\\
I-2&4500&1.10&1.45&-1.36&\nodata&-0.57&-0.57\\
I-50&4525&1.15&1.40&-1.35&\nodata&-0.54&-0.54\\
II-74&4525&1.30&1.30&-1.25&\nodata&-0.67&-0.67\\
III-59&4575&1.20&1.35&-1.31&\nodata&-0.53&-0.53\\
IV-36&4575&1.50&1.35&-1.30&\nodata&-0.63&-0.63\\
II-50&4590&1.57&1.45&-1.20&\nodata&-0.66&-0.66\\
IV-30&4625&1.00&1.75&-1.44&\nodata&-0.63&-0.63\\
III-52&4625&1.50&1.45&-1.34&\nodata&-0.54&-0.54\\
IV-4&4625&1.55&1.20&-1.25&\nodata&-0.70\tablenotemark{a}&-0.70\tablenotemark{a}\\
IV-26&4650&1.05&1.40&-1.40&\nodata&-0.55&-0.55\\
III-53&4700&1.05&1.75&-1.46&\nodata&-0.63&-0.63\\
I-55&4700&0.85&1.80&-1.48&\nodata&-0.57&-0.57\\
\sidehead{NGC 288  \citep{Shetrone2000}}                        
531&3780&0.10&1.60&-1.31&\nodata&-0.50\tablenotemark{a}&-0.50\tablenotemark{a}\\
403&3950&0.20&1.90&-1.43&\nodata&-0.37&-0.37\\
274&4025&0.70&1.90&-1.37&\nodata&-0.33&-0.33\\
344&4180&0.80&1.60&-1.36&\nodata&-0.68&-0.68\\
245&4250&0.80&1.40&-1.41&-0.10&-0.40&-0.40\\
231&4300&1.10&1.50&-1.41&-0.25&-0.32&-0.29\\
388&4325&1.30&1.60&-1.37&\nodata&-0.32&-0.32\\
297&4330&1.20&1.70&-1.41&-0.30&-0.32&-0.31\\
351&4330&1.20&1.55&-1.33&-0.30&-0.48&-0.39\\
287&4350&1.20&1.40&-1.45&\nodata&-0.32&-0.32\\
307&4350&1.20&1.35&-1.40&\nodata&-0.45&-0.45\\
\sidehead{M3  \citep{Kraft1992}}
VZ-1397&3950&0.40&2.35&-1.46&\nodata&-0.50&-0.50\\
AA&4000&0.40&2.25&-1.45&\nodata&-0.55&-0.55\\
II-46&4000&0.60&2.10&-1.46&\nodata&-0.65&-0.65\\
VZ-297&4070&0.70&2.25&-1.52&\nodata&-0.55&-0.55\\
III-28&4160&0.75&1.75&-1.49&\nodata&-0.65&-0.65\\
VZ-1000&4175&0.45&2.10&-1.45&\nodata&-0.50&-0.50\\
VZ-1127&4225&0.90&2.00&-1.48&\nodata&-0.65&-0.65\\
\sidehead{M10  \citep{Kraft1995}}
A-II-24&4050&0.10&2.00&-1.50&\nodata&-0.69&-0.69\\
A-III-21&4060&0.50&2.10&-1.49&\nodata&-0.50&-0.50\\
H-I-367&4135&0.60&1.70&-1.50&\nodata&-0.72&-0.72\\
A-III-16&4150&0.90&2.00&-1.50&\nodata&-0.47\tablenotemark{a}&-0.47\tablenotemark{a}\\
B&4150&0.50&1.80&-1.49&\nodata&-0.64&-0.64\\
C&4200&0.75&2.00&-1.62&\nodata&-0.60&-0.60\\
D&4200&1.05&2.00&-1.49&\nodata&-0.55&-0.55\\
H-I-15&4225&0.75&1.75&-1.50&\nodata&-0.65&-0.65\\
E&4350&0.80&2.00&-1.60&\nodata&-0.70&-0.70\\
A-I-60&4400&1.10&1.60&-1.50&\nodata&-0.65&-0.65\\
A-III-5&4400&1.20&1.75&-1.36&\nodata&-0.60&-0.60\\
A-I-61&4550&1.00&2.00&-1.66&\nodata&-0.70\tablenotemark{a}&-0.70\tablenotemark{a}\\
G&4650&1.20&1.85&-1.56&\nodata&-0.85\tablenotemark{a}&-0.85\tablenotemark{a}\\
\sidehead{NGC 7006  \citep{Kraft1998}}
I-1&3900&0.10&2.25&-1.50&\nodata&-0.65&-0.65\\
V19&4100&0.3&2.4&-1.57&\nodata&-0.65&-0.65\\
II-103&4200&0.75&1.85&-1.54&\nodata&-0.50&-0.50\\
II-46&4200&0.5&1.75&-1.56&\nodata&-0.55&-0.55\\
\sidehead{M13  \citep{Kraft1997}}
L-598&3900&0.00&2.10&-1.62&-0.57&-0.55&-0.56\\
I-48&3920&0.30&2.00&-1.57&-0.60&-0.67&-0.64\\
L-629&3950&0.20&2.00&-1.64&-0.75&-0.58&-0.67\\
II-67&3950&0.20&2.10&-1.52&-0.67&-0.67&-0.67\\
II-90&4000&0.30&2.00&-1.58&-0.50&-0.65&-0.58\\
IV-25&4000&0.15&2.25&-1.56&-0.45&-0.75&-0.60\\
L-835&4090&0.55&1.90&-1.52&\nodata&-0.65&-0.65\\
I-12&4600&1.50&1.60&-1.58&\nodata&-0.60&-0.60\\
IV-19&4650&1.50&1.60&-1.60&\nodata&-0.60&-0.60\\
II-09&4700&1.70&1.50&-1.58&\nodata&-0.58&-0.58\\
IV-22&4700&1.90&1.50&-1.54&\nodata&-0.60&-0.60\\
II-41&4750&2.00&1.75&-1.48&\nodata&-0.65&-0.65\\
I-72&4850&1.90&1.40&-1.64&\nodata&-0.40\tablenotemark{b}&-0.40\tablenotemark{b}\\
II-01&4850&2.10&1.25&-1.56&\nodata&-0.60&-0.60\\
II-28&4850&1.75&2.00&-1.64&\nodata&-0.65\tablenotemark{a}&-0.65\tablenotemark{a}\\
I-54&4975&1.70&1.75&-1.68&\nodata&-0.55\tablenotemark{b}&-0.55\tablenotemark{b}\\
\sidehead{M15  \citep{Sneden1997}}
K386&4200&0.15&1.85&-2.38&-0.65&\nodata&-0.65\\
K462&4225&0.30&1.85&-2.43&-0.70&\nodata&-0.70\\
K634&4225&0.30&1.85&-2.42&-0.57&\nodata&-0.57\\
K583&4275&0.30&1.90&-2.42&-0.66&\nodata&-0.66\\
K853&4275&0.50&1.85&-2.38&-0.80&\nodata&-0.80\\
K702&4325&0.25&1.90&-2.42&-0.73&\nodata&-0.73\\
K479&4325&0.45&2.30&-2.42&-0.70\tablenotemark{b}&\nodata&-0.70\tablenotemark{b}\\
K490&4350&0.60&1.65&-2.44&-0.45&\nodata&-0.45\\

\tablenotetext{a}{error =$\pm$0.2 dex}
\tablenotetext{b}{upper limit only}
\enddata
\end{deluxetable}

\clearpage

\begin{deluxetable}{ccccccccc}
\tabletypesize{\scriptsize}
\tablecaption{Cluster average [Cu/Fe] \label{tbl-2}}
\tablewidth{0pt}
\tablehead{
\colhead{Cluster}  & 
\colhead{[Fe/H]$_{avg}$}   & \colhead{[Cu/Fe]$_{avg}$} & \colhead{$\sigma$}& \colhead{[Fe/H]$_{avg}$} & \colhead{[Cu/Fe]$_{avg}$} & \colhead{[Fe/H]$_{avg}$} & \colhead{[Cu/Fe]$_{avg}$}\\

\colhead{}  & 
\colhead{MARCS}   & \colhead{MARCS} & \colhead{}& \colhead{Kurucz} & \colhead{Kurucz} & \colhead{corrected} & \colhead{corrected}
}
\startdata

M71           &-0.79&-0.17&0.05&-0.72&-0.24&-0.77&-0.19\\
M4             &-1.18&-0.32&0.06&-1.18&-0.32&-1.23&-0.27\\
NGC 362   &-1.33&-0.52&0.06&-1.33&-0.52&-1.38&-0.47\\
M5             &-1.35&-0.58&0.08&-1.35&-0.58&-1.40&-0.53\\
NGC 288   &-1.39&-0.40&0.11&-1.39&-0.40&-1.44&-0.35\\
M3            &-1.46&-0.59&0.07&-1.45&-0.60&-1.50&-0.55\\
M10          &-1.52&-0.64&0.10&-1.51&-0.65&-1.56&-0.60\\
NGC 7006 &-1.55&-0.59&0.08&-1.54&-0.60&-1.59&-0.55\\
M13          &-1.58&-0.60&0.06&-1.57&-0.61&-1.62&-0.56\\
M15          &-2.41&-0.66&0.11&-2.31&-0.76&-2.36&-0.71\\

\enddata
\end{deluxetable}

\clearpage

\end{document}